\documentclass[12pt]{article}
\usepackage{amssymb, euscript}
\usepackage{graphicx}
\usepackage[hang,small,bf]{caption}

\newcommand{\be}[3]{\begin{equation}  \label{#1#2#3}}
\newcommand{\ee}{\end{equation}}
\newcommand{\ba}{\begin{array}}
\newcommand{\ea}{\end{array}}

\begin{document}

%%%%%%%%%%%%%%%%%%%%%%%%%%%%%%%%%%%%%%%%%%%%%%%%%%%%%%

%%%%%%%%%%%%%%%%%%%%%%%%%%%%%%%%%%%%%%%%%%%%%%%%%%%

\begin{flushright}
\hfill{AEI-2002-028}\\
\hfill{hep-th/0204061}

\end{flushright}

\vspace{15pt}

\begin{center}{ \Large{\bf 
Singular 7-manifolds with $G_2$ holonomy \\[5mm]
and intersecting 6-branes
}}

\vspace{30pt}

{
{\bf Klaus Behrndt}\footnote{E-mail: behrndt@aei.mpg.de} 
 }
\vspace{15pt}

{\it Max-Plank-Institut f\"ur Gravitationsphysik \\
Albert Einstein Institut\\ 
Am M\"uhlenberg 1,  14476 Golm,
 Germany}
\vspace{80pt}   
  
{ABSTRACT}
\end{center} 

A 7-manifold with $G_2$ holonomy can be constructed as a $\mathbb
R^3$ bundle over a quaternionic space. We consider a quaternionic base
space which is singular and its metric depends on three parameters,
where one of them corresponds to an interpolation between ${\mathbb
S}^4$ and ${\mathbb C \mathbb P}^2$ or its non-compact analogs. This
4-d Einstein space has four isometries and the fixed point set of a
generic Killing vector is discussed. When embedded into M-theory the
compactification over a given Killing vector gives intersecting
6-branes as IIA configuration and we argue that membrane instantons
may resolve the curvature singularity.

\newpage

%%%%%%%%%%%%%%%%%%%%%%%%%%%%%%%%%%%%%%%%%%%%%

\section{Introduction}

%%%%%%%%%%%%%%%%%%%%%%%%%%%%%%%%%%%%%%%%%%%%%%%

Minimal supersymmetric field theories in four dimensions have many
phenomenologically interesting features and can be obtained by
Calabi-Yau compactification of heterotic string theory. Over the years
this has been the standard way for their construction, because one
obtains naturally chiral fermions and non-Abelian gauge groups. On the
other hand, it is much more difficult to obtain phenomenological
interesting $N=1$ models directly from M-theory, see \cite{420,
370}. One reason is that the 7-manifold has to have $G_2$-holonomy
and these spaces are not yet well understood. Only few non-compact
examples, that rely on the construction done in \cite{110, 100}, are
explicitly known.  Another reason are the difficulties to obtain a
model with chiral fermions in four dimensions \cite{640}. In fact,
chiral fermions, but also non-Abelian gauge groups, require that the
7-manifold is singular \cite{650, 590, 350, 360} and a supergravity
approximation may become questionable.

There are mainly two classes of known metrics: one is topologically a
$\mathbb R^3$ bundle over a quaternionic space and the other a
$\mathbb R^4$ bundle over $\mathbb S^3$. They have been introduced in
\cite{110,100} and many generalizations, with more parameters or
functions, have been discussed in the past year. It is impossible to
give a complete list of refs., but relevant aspects for our
consideration can be found in \cite{550, 240, 270, 360, 230, 250, 530,
540, 660, 720}.  In the first class one can further distinguish between
different quaternionic spaces as e.g.\ the 4-sphere ${\mathbb
S}^4={SO(5) \over SO(4)}$ but also the complex projective space
${\mathbb C \mathbb P}^2 = {SU(3) \over U(2)}$, which are the only
compact homogeneous quaternionic 4-dimensional spaces \cite{320}.
Apart from their non-compact analogs, there are also non-homogeneous
quaternionic spaces, which we want to use for the construction of
explicit metrics.

Quaternionic spaces have been discussed in the physical literature
mainly as moduli spaces of $N=2$ supergravity in four and five
dimensions \cite{620, 610}, for a more recent discussion see
also \cite{630,490}. The quantum moduli space is expected to be
non-homogeneous, but unfortunately not many examples are explicitly
known. In a long quest to incorporate brane world scenarios into
gauged supergravity and to overcome numerous problems, see e.g.\
\cite{670,680}, a non-homogeneous quaternionic space has been explored
recently in \cite{150}. This space interpolates between the two
homogeneous spaces and it is obvious to consider this space also in
the construction of metrics with $G_2$ holonomy and therewith ''to
unify'' the two spaces representing $\mathbb R^3$ bundles over
$\mathbb S^4$ and $\mathbb{CP}^2$. More important is however, that the
Killing vectors of this non-homogeneous space have additional fixed
points which become additional D6-branes after compactification along
this Killing vector \cite{170}. To be more clear, a co-dimension four
fixed point set of a given isometry extends in 6+1 dimensions
representing a NUT (=point) on the quaternionic space can be
identified as a 6-brane. There are also co-dimension two fixed points,
but the interpretation of these bolts on the quaternionic space is
unclear.

As it has been shown in \cite{500} on any component of a non-compact
{\em homogeneous} quaternionic spaces can be at most one fixed point
and for compact we expect at most two. Hence, in order to find a
configuration with three or more 6-branes one necessarily needs in
this setup non-homogeneous quaternionic spaces. As we will see,
depending on the choice of parameters, our space can have up to
five NUT fixed points. The additional fixed points reflect the
topological non-trivial nature of this manifold, because, due to the
Lefschetz fixed point theorem, every NUT fixed point adds one unit to
the Euler characteristic of the manifold, see also \cite{360, 250}.

The paper is organized as follows. In the next section we basically
follow the literature and solve the Killing spinor equations followed
by a discussion of the closed and co-closed 3-form. This consideration
is very general without using any specific quaternionic space. In
section 3 we will investigate examples corresponding to different
quaternionic spaces. In section 3.1 we start with the homogenous cases
and in section 3.2 we discuss the following basic features of the
non-homogenous space: it has four Killing vectors; it exhibits a
curvature singularity, which separates two asymptotic regions and
finally it interpolates between the two homogeneous spaces. In section
4 we give a detailed analysis of the isometries and the fixed point
set; many details for the non-compact case were already derived in
\cite{150}.  Due to their interpretation as D6-branes, we especially
identify NUT fixed points.  For compact quaternionic spaces there can
be three or five and for non-compact we found three or four
non-degenerate fixed points, depending on the isometry groups $U(1)
\times SU(2)$ or $U(1) \times SL(2,R)$. Unfortunately, in the IIA
description the string coupling constant is not bounded -- every model
contains at least one non-compact direction in which the dilaton
diverges. We conclude with a discussion of two aspects that could be
interesting for future investigations: (i) resolution of the curvature
singularity e.g.\ by of membrane instantons (or general $G$-fluxes)
and (ii) the construction of new Spin(7) manifold in complete analogy
to the procedure of this paper.

%%%%%%%%%%%%%%%%%%%%%%%%%%%%%%%%%%%%%%%%%%%%

\section{Constructing metrics with $G_2$ holonomy}

%%%%%%%%%%%%%%%%%%%%%%%%%%%%%%%%%%%%%%%%%%%%%

Before we can discuss explicit examples let us summarize some aspects
of the procedure described in \cite{110,130,120} which will also fix
our notations. After discussing the ansatz for the metric, we will
derive the Killing spinor and the closed and co-closed 3-form. Both
conditions are sufficient to ensure supersymmetry for this background.

%%%%%%%%%%%%%%%%%%%%%%%%%%%%%%%%%%%%%%%%%%%

\subsection{The metric ansatz}

%%%%%%%%%%%%%%%%%%%%%%%%%%%%%%%%%%%%%%%%%%%

The construction of the 7-manifold relies on a 4-d quaternionic base
space and before we discuss the metric ansatz we need some basic
properties of these spaces; for a recent resume about quaternionic
geometry we refer to \cite{630} and the appendix of \cite{490}.
Quaternionic spaces are generalizations of complex spaces that allow
for three complex structures $J^i$ ($i = 1,2,3$) defined by the
algebra
\be130
J^i \cdot J^j = - {\mathbb I} \, \delta^{ij} + \epsilon^{ijk} J^k 
\ee
and denoting the quaternionic vielbein by $e^m$, one obtains three 2-forms
$\Omega^i$ by
\be150
\Omega^i = e^m \wedge J^i_{mn} e^n \ .
\ee
The holonomy of a 4n-dimensional quaternionic spaces is contained in
$Sp(n) \times SU(2)$. This statement is trivial for $n=1$ and is replaced
by the requirement that the Weyl-tensor of 4-dimensional quaternionic
space has to be anti-selfdual \cite{620}
\[
W + {\, ^{\star}W} = 0 \ .
\]
For a quaternionic space in any dimensions the triplet of 2-forms
$\Omega^i$ is expressed in terms of the $SU(2)$-part of the
quaternionic connection $A^i$ as
\be120
dA^i + {1 \over 2} \, \epsilon^{ijk} A^j \wedge A^k = 
	{\kappa} \, \Omega^i 
\ee
which ensures that the triplet of 2-forms is covariantly constant.
Moreover, any quaternionic space is an Einstein space with the
curvature $\kappa$ implying that its metric $g_{mn}$ solves the
equation
\be170
R_{mn} = 3 \, \kappa \, g_{mn} \ .
\ee
The complex structures can be selfdual or anti-selfdual and in our
notation we will take the latter one ($J^i_{mn} = -{1 \over 2}
\epsilon_{mnpq} \, J^i_{pq}$) so that the triplet of 2-forms can be
written as
\be160
\ba{l}
\Omega^1 = e^4 \wedge e^7 - e^5 \wedge e^6\ , \\[2mm]
\Omega^2 = e^4 \wedge e^6 + e^5 \wedge e^7 \ , \\[2mm]
\Omega^3 = -e^4 \wedge e^5 + e^6 \wedge e^7 \ .
\ea
\ee
Moreover, the $SU(2)$ connection is given as the anti-selfdual part
of the spin connection $\omega^{mn}$ of the quaternionic space
\be190
A^i = {1 \over 2} \omega^{mn} J_{mn}^i \ .
\ee
In the same way, the selfdual part gives the $Sp(n)$ connection.

Having the basic relations for the quaternionic base space, the metric
of the 7-manifold is introduced by the ansatz \cite{110, 100}
\be100
\hat{ds}^2 = e^{2f} \alpha^i \alpha^i + e^{2g} e^m e^m
\ee
where $e^m$ ($m=4...7$) is the vielbein of the quaternionic space
and $\alpha^i$ ($i = 1..3$) is defined by
\be110
\alpha^i \equiv \nabla u^i = d u^i + \epsilon^{ijk} A^j u^k 
\ee
with $u^i$ as local coordinates (in addition to the quaternionic once).
Using the relation (\ref{120}), it is straightforward to verify that 
\be122
\nabla \alpha^i \equiv d\alpha^i + \epsilon^{ijk} A^j \alpha^k =
	{\kappa} \, \epsilon^{ijk} \, \Omega^j \, u^k \ .
\ee
The two unknown functions $f$ and $g$ in the metric ansatz are now
fixed by the requirement of $G_2$ holonomy implying the existence of a
Killing spinor or equivalently the existence of a closed and co-closed
3-form. We will check both conditions.

%%%%%%%%%%%%%%%%%%%%%%%%%%%%%%%%%%%%%%%%%%%%%%%%%%%

\subsection{Solving the Killing spinor equations}

%%%%%%%%%%%%%%%%%%%%%%%%%%%%%%%%%%%%%%%%%%%%%%%%%%%%%

\begin{figure} 
\begin{center}
\includegraphics[angle=0,width=50mm]{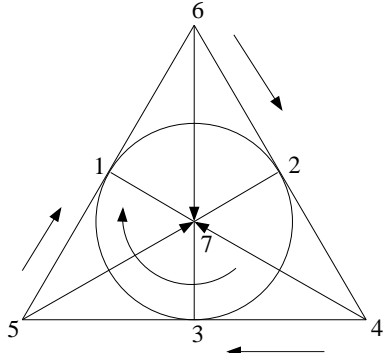}
\end{center}
\caption{The multiplication table of imaginary components of
octonians: $i_1 i_2 = i_3$, $i_6 i_2 = i_4$, $i_4 i_7 = i_1$, ...\ can
be obtained from this figure by following the lines in the direction of
the arrows.}
\label{fig1}
\end{figure}

Consider M-theory on the manifold $M_4 \times X_7$ and assume that
$M_4$ is the flat 4-d Minkowski space. If we moreover assume the
absence of G-fluxes, a Killing spinor $\epsilon$ is a solution of the
equation ($a,b = 1...7$)
\be200
(d + {1 \over 4} \, \hat \omega^{ab} \Gamma_{ab}) \, \epsilon  = 0 
\ee
where $\hat \omega^{ab}$ is the spin connection 1-form and
$\Gamma_{ab} = \Gamma_{[a} \Gamma_{b]}$ with $\Gamma_a$ as the 7-d
gamma matrices.  If this equation has exactly one solution, the
resulting 4-dimensional field theory has $N=1$ supersymmetry and the
manifold $X_7$ has $G_2$ holonomy. Recall, the unrestricted holonomy
of a 7-manifold is $SO(7)$, but the existence of a (covariantly
constant) Killing spinor implies that the holonomy of the manifold is
restricted. A generic spinor transforms as representation ${\bf 8}$ of
$SO(7)$, while a Killing spinor implies the decomposition ${\bf
8 \rightarrow 7+1}$, which is exactly the decomposition under $G_2
\subset SO(7)$. The exceptional group $G_2$ appears as automorphism
group of octonions: $o = x^0 {\mathbb I} + x^a i_a$, where $i_a$
satisfy the algebra 
\[
i_a i_b = -\delta_{ab} + \psi_{abc} \, i_c \ ,
\]
for more details see e.g.\ in \cite{140, 130, 400, 410}.  The $G_2$-invariant
3-index tensor $\psi_{abc}$ can be obtained from figure \ref{fig1} and is
given in the standard basis by
\be230
\ba{rcl}
{1 \over 3!} \psi_{abc} \, e^a \wedge e^b \wedge e^c &=&
e^1 \wedge e^2 \wedge e^3 + e^4 \wedge e^3 \wedge e^5 +
e^5 \wedge e^1 \wedge e^6 + e^6 \wedge e^2 \wedge e^4 + \\[2mm] && 
e^4 \wedge e^7 \wedge e^1 +e^5 \wedge e^7 \wedge e^2 +
e^6 \wedge e^7 \wedge e^3 \ , \\[2mm] 
&=& e^1 \wedge e^2 \wedge e^3 + e^i \wedge \Omega^i
\ea
\ee
where we used the complex structures as introduced in (\ref{150}) and
(\ref{160}).

If the Killing spinor equation (\ref{200}) has more than one solution,
the holonomy is smaller than $G_2$ resulting in a 4-d field theory
with $N>1$ supersymmetry. The holonomy is equal to $G_2$ if the
Killing spinor equation has exactly one solution, which is equivalent
to the absence of covariantly constant 1-forms \cite{110}. The
existence of such a 1-form would imply the existence of a covariantly
constant Killing vector and hence to a factorization of the space. As
we will discuss in the next section this is not the case for the
examples that we consider and hence our models yield $N=1$
supersymmetry in 4 dimensions.

Following the arguments from \cite{130,120} (see also \cite{700,710}),
one can define two orthogonal projectors ${\mathbb P_{\pm}}$ that
decompose the spin connection into two parts: $\hat\omega^{ab} =
\hat\omega_+^{ab} + \hat\omega_-^{ab}$ with
\be270
\ba{l}
\hat\omega^{ab}_+ = {\mathbb P}_{+\ cd}^{ab}\, \hat\omega^{cd} \equiv
{2 \over 3} (\hat \omega^{ab} + {1 \over 4} \psi^{abcd} \hat\omega_{cd}) \  , 
\\[3mm]
\hat\omega^{ab}_- = {\mathbb P}_{-\ cd}^{ab}\, \hat\omega^{cd} \equiv
{1 \over 3} (\hat \omega^{ab} - {1 \over 2} \psi^{abcd} \hat\omega_{cd}) 
\ea
\ee
where $\psi_{abcd}$ is the $G_2$ invariant 4-index tensor which is
dual to the 3-index tensor ($\psi_{abcd} = {1 \over 3!}
\epsilon_{abcdefg} \, \psi_{efg}$). In order to solve (\ref{200}) one
imposes on the (constant) Killing spinor $\epsilon$ and the spin
connection $\hat \omega^{cd}$ the projector equations
\be210
{\mathbb P}_{+\ cd}^{ab}\, \Gamma^{cd} \epsilon = 0 \quad ,\qquad
{\mathbb P}_{-\ cd}^{ab}\, \hat\omega^{cd} = 0 
\ee
and finds as solution for the Killing spinor \cite{130}
\be290
\epsilon^\alpha = c \, \delta^{\alpha 8}
\ee
where $\alpha = 1 \ldots 8$ is the $SO(7)$ spinor index and $c$ is
constant. The projector condition on the spin connection can be
simplified by contracting with the 3-index tensor
$\psi_{abc}$ and using the relation: $\psi_{abc} \psi_{bcde} = - 4
\psi_{abc}$. One infers \cite{120}
\be220
\psi_{abc} \, \hat \omega^{bc} = 0  
\ee
yielding first order differential equations for the unknown
functions. We will assume in the following that in the metric ansatz
(\ref{100}) $f = f(|u|)$ and $g=g(|u|)$ and obtain for the spin
connection
\be240
\ba{rcl}
\hat \omega^{ij} &=&  {f' \over |u|} \, \alpha^{[i} u^{j]} - \epsilon^{ijk}
	A^k \ , \\[2mm]
\hat \omega^{mn} &=& \omega^{mn} - {\kappa} e^{2(f-g)}
	\epsilon^{ijk} u^k J^j_{mn} \alpha^i \ , \\[2mm]
\hat \omega^{mi} &=& {g' \over |u|} \, e^{g-f} \, e^m u^i - {\kappa}
	e^{f -g} \, \epsilon^{ijk} \, u^k J^j_{mn} e^n 
\ea
\ee
where $(...)'$ denotes the derivative with respect to $|u|$. In the
contraction with the 3-index tensor (\ref{230}) we use the relation
(\ref{190}) and find
\be250
\ba{l}
0 = \psi_{iab} \, \hat \omega^{ab} = \psi_{ijk} \hat \omega^{jk} 
	+ \psi_{imn} \hat \omega^{mn} = \Big[{f' \over |u|} + 
	\kappa\, e^{2(f-g)}\Big] 
\epsilon^{ijk} \, \alpha^j u^k \ , \\[2mm]
0 = \psi_{mab} \, \hat \omega^{ab} = \psi_{mni}\,  \hat \omega^{ni} 
	= J^i_{mn} \hat \omega^{ni} = 
	\Big[ {g' \over |u|}  - \kappa \, e^{2(f-g)}\Big]
	e^{g-f}\, J^i_{mn} e^n \, u^i \ .
\ea
\ee
These equations have the symmetry: $f \rightarrow f + \lambda$, $g
\rightarrow g + \lambda$, giving a constant conformal rescaling of the
metric (\ref{100}) and a second symmetry: $f \rightarrow f - \lambda$,
$u \rightarrow e^{\lambda} \, u$ leaves the metric invariant. Using
these symmetries, only one (discrete) integration constant $c = \pm 1
,0$ appears in the solution that can be written as \cite{110}
\be260
e^{-4f} = 2 \kappa |u|^2 + c \qquad , \qquad 
e^{4g} = 2 \kappa |u|^2 + c
\ee
with $|u|^2 = (u^1)^2 + (u^2)^2 + (u^3)^2$.  Notice if $\kappa <0$
(i.e.\ a non-compact quaternionic space), the parameter range of $u$
is bounded by $2 \kappa |u|^2 + c > 0$, while for $\kappa \geq 0$ $u$ is
unbounded.

%%%%%%%%%%%%%%%%%%%%%%%%%%%%%%%%%%%%%%%%%%%

\subsection{The closed and co-closed 3-form}

%%%%%%%%%%%%%%%%%%%%%%%%%%%%%%%%%%%%%%%%%%%

The holonomy reduction from $SO(7)$ to $G_2$ is equivalent to the
existence a 3-form $\Phi$ which is closed and co-closed. Again
following the procedure done in \cite{110} we write this 3-form as
\be320
\Phi = {1 \over 3!} \psi_{abc} \, \hat e^a \wedge \hat e^b \wedge \hat e^c=
e^{3f} \, \alpha^1 \wedge \alpha^2 
	\wedge \alpha^3 + e^{f+2g} \, \alpha^i \wedge \Omega^i  
\ee
and we will show now that the functions (\ref{260}) ensure that:
$d\Phi = d\, {^{\star}\Phi} = 0$, i.e.\ it is closed and co-closed.
Since $d(u^i u^i) = d|u|^2 = 2|u|d|u| = 2 u^i \alpha^i$ and $\nabla
\Omega^i =0$ as well as using the relation (\ref{122}) it follows
\be340
\ba{rcl}
{1\over 3!} \, d(\epsilon_{ijk} \alpha^i \wedge \alpha^j \wedge \alpha^k) 
	&=& 
	{1 \over 2} \epsilon_{ijk} (\nabla \alpha^i) \wedge \alpha^j \wedge
	\alpha^k = - {\kappa}\, |u| d|u| \wedge \alpha^i \wedge \Omega^i
	\ , \\[2mm] 
d(\alpha^i \wedge \Omega^i) &=& (\nabla \alpha^i) \wedge \Omega^i - 
	\alpha^i (\nabla \Omega^i) = 0 \ , \\[2mm]
(d e^{3f}) \wedge \alpha^1 \wedge \alpha^2 \wedge \alpha^3 &=&
\big(e^{3f}\big)' \, d|u| \wedge   \alpha^1 \wedge \alpha^2 \wedge \alpha^3 
= 0 \ .
\ea
\ee
Therefore, from $d\Phi=0$ we derive the equation
\be350
{\kappa}  \, |u| \, e^{3f} =  (e^{f + 2g})' \ .
\ee
Next, for the co-closure we have to check whether  the dual
4-form
\be310
\Psi = e^{2(f+g)}\, {1 \over 2} \epsilon_{ijk} \, 
	\alpha^i \wedge \alpha^j \wedge 
	\Omega^k +  e^{4g} \, e^4 \wedge e^5 \wedge e^6 \wedge e^7 
\ee
is also closed.  Employing the relation $\Omega^i \wedge \Omega^j = -
2 \, \delta^{ij}\, e^4 \wedge e^5 \wedge e^6 \wedge e^7$ and again
(\ref{122}) one obtains
\be360
{1 \over 2} d (\epsilon_{ijk} \alpha^i \wedge \alpha^j \wedge \Omega^k)
= \epsilon_{ijk} (\nabla \alpha^i)\wedge \alpha^j \wedge \Omega^k
= -4 \kappa \, u^i \alpha^i  \wedge e^4 \wedge e^5 \wedge e^6 \wedge e^7 
\ee
and thus (recall $u^i \alpha^i = |u| d |u|$)
\be370
\ba{rcl}
0 = d\Psi &=& \big(e^{2(f+g)} \big)' d|u| \wedge {1 \over 2} \epsilon_{ijk}
	\alpha^i \wedge \alpha^j \wedge \Omega^k +\\[2mm]
	&& \Big[-4 \kappa |u| \, e^{2(f+g)} + \big(e^{4g} \big)' \Big]
	d|u| \wedge e^4 \wedge e^5 \wedge e^6 \wedge e^7 \ .
\ea
\ee
This gives the two differential equations 
\be380
\big(\, e^{2(f+g)}\, \big)' = 0 \qquad , \qquad 
\big( e^{4g} \big)' = 4\kappa |u| \, e^{2(f+g)} 
\ee
and it is straightforward to verify that these two equations together
with (\ref{350}) are equivalent to the first order equations
derived from (\ref{250}) with the solution (\ref{260}).

%%%%%%%%%%%%%%%%%%%%%%%%%%%%%%%%%%%%%%%%%%%%%%%%%

\section{Explicit metric of the 7-manifold}

%%%%%%%%%%%%%%%%%%%%%%%%%%%%%%%%%%%%%%%%%%%%%%%%%%

In the previous section we have verified that the metric of the
7-manifold with $G_2$ holonomy is given by \cite{110,100}
\be510
ds^2 = {1 \over \sqrt{2 \kappa |u|^2 + c}}\,  
\big( du^i + \epsilon^{ijk} A^j u^k \big)^2
+ \sqrt{2 \kappa |u|^2 + c} \; e^m e^m \ , 
\ee
where $e^m$ is the vielbein of the quaternionic space with the
$SU(2)$ connection $A^j$ as introduced in (\ref{190}).  By choosing
different quaternionic spaces we can now discuss explicit models.

%%%%%%%%%%%%%%%%%%%%%%%%%%%%%%%%%%%%%%%%%%%%%%

\subsection{Homogeneous quaternionic spaces}

%%%%%%%%%%%%%%%%%%%%%%%%%%%%%%%%%%%%%%%%%%%%%

We will start with examples that played an important role in the recent
literature.  They base on homogeneous quaternionic spaces, which
appear in two classes \cite{320,330}, namely

\medskip

(i) the maximal symmetric spaces ${SO(5) \over SO(4)}$ and ${SO(4,1)
\over SO(4)}$ with 10 isometries and

(ii) the complex projective spaces ${SU(3) \over SU(2)\times U(1)}$ 
and ${SU(2,1) \over SU(2)\times U(1)}$ each having 8 isometries. 

\medskip

\noindent
The metric of the quaternionic space and the corresponding $SU(2)$
connection for the maximal symmetric case (i) can be written as
\be520
\ba{l}
e^m e^m = {d\rho^2 \over 1 - \kappa \rho^2} + (1 - \kappa \rho^2) \, d\psi^2
	+ \rho^2 (d\theta^2 + \sin^2 \theta d \varphi^2 ) \ , \\[3mm]
A^1 = - \sqrt{1 - \kappa \rho^2} \, \sin\theta \, d \varphi \ , \quad 
A^2 =- \sqrt{1 - \kappa \rho^2} \, d\theta \ , \\[3mm]
A^3 = - \kappa \rho \, d\psi - \cos\theta \, d \varphi \ .
\ea
\ee
For $\kappa = +1$ this metric describes the coset ${SO(5) \over SO(4)}
= {\mathbb S^4}$, and for $\kappa = -1$ the corresponding non-compact
analog, ${SO(4,1) \over SO(4)}$. These spaces with the maximal number
of Killing vectors become trivial for $\kappa=0$ and have a vanishing
Weyl tensor. The complex projective spaces (ii) have less isometries
and the Weyl tensor is non-trivial.  The corresponding expressions
read (cp.\ \cite{690})
\be530
\ba{l}
e^m e^m = {2\, d\rho ^2 \over (1+ \kappa \rho^2)^2} + {\rho^2 \over 2
	(1+ \kappa \rho^2)^2}\, (d\psi - \cos\theta \, d \varphi)^2 + 
	{\rho^2 \over 2 (1 + \kappa \rho^2)} \big(
	d\theta^2 + \sin^2\theta \, d\varphi^2
	\big) \\[3mm]
A^1 = -{\sin\theta \over \sqrt{1 + \kappa \rho^2}} \, d\varphi\ , \quad 
A^2 = -{1 \over \sqrt{1 + \kappa \rho^2}} \, d\theta\  , \\[3mm] 
A^3 = -{\kappa \rho^2 \over 2(1 + \kappa \rho^2)} \, d\psi 	
	- {(2+\kappa \rho^2) \cos\theta \over 2(1 + \kappa \rho^2)}
	\, d\varphi \ .
\ea
\ee
As before, the space is compact for $\kappa>0$ and otherwise
non-compact.

Both spaces are spherical symmetric related to the ${\mathbb S^2}$
parameterized by $(\theta , \varphi)$. But there is also a
generalization to any 2-space with constant curvature $\epsilon$ (as
it will appear in the next example). If we take into account this
additional parameter, the resulting $G_2$ metric (\ref{510}) depends
in total on three parameters: $\epsilon$, $c$ and $\kappa$.

%%%%%%%%%%%%%%%%%%%%%%%%%%%%%%%%%%%%%%%%%%%%%%%%%%%%%%%%

\subsection{Quaternionic spaces with four isometries }

%%%%%%%%%%%%%%%%%%%%%%%%%%%%%%%%%%%%%%%%%%%%%%%%%%%%%%

More general, non-homogeneous quaternionic spaces may be classified by
the number of isometries, but not many concrete examples are known;
see however \cite{510, 520}. One example with four isometries has been
discussed in \cite{150} and when regarded as a hyper multiplet moduli
space yields upon gauging a supersymmetric Randall-Sundrum
scenario. The metric and connection for this space is given by
\be540
\ba{l}
e^m e^m = {d\rho ^2 \over V(\rho)} + V(\rho) 
	\Big[d\tau  - n\, {y dx- x dy \over 1 + {\epsilon \over 4} (x^2 + y^2)}
	\Big]^2 + (\rho^2 - n^2) {dx^2 + dy^2 \over \big[1 + {\epsilon \over 4}
 	(x^2 + y^2)\big]^2} \\[3mm]
A^1 = - \sqrt{{\rho + n \over \rho -n} \, V} \, {dy \over
	1+ {\epsilon \over 4 }(x^2 + y^2)} \ , \quad  
A^2 = -\sqrt{{\rho + n \over \rho -n} \, V} \, {dx \over
	1+ {\epsilon \over 4 }(x^2 + y^2)}\   , \\[3mm]
A^3 = -\kappa(\rho -n) \, d\tau - \Big[\epsilon + 2n\kappa(n - \rho)
	\Big]\, {ydx - xdy
	\over 1 +{\epsilon \over 4} (x^2 + y^2)} \ , 
\ea
\ee
with 
\be541
V = -\kappa \, {(\rho -n)
	 \over \rho+n} \, (\rho-\rho_+)(\rho-\rho_-) \ , \quad
\rho_{\pm} = - n \pm\sqrt{4n^2 + {\epsilon \over \kappa}} \ . 
\ee
As we mentioned in the last paragraph the parameter $\epsilon=0, \pm
1$ determines the symmetry group: for $\epsilon = +1$ it is a spherical
symmetric solution with the isometry group $U(1) \times SU(2)$; for
$\epsilon = -1$ the isometry group becomes $U(1) \times SL(2,R)$ and for
$\epsilon =0$ it is the solvable sub-algebra of $SO(1,4)$.

To understand this space better consider $\epsilon = 1$, where it
becomes the metric of the Taub-NUT-(A)dS space \cite{450}, with $n$
as NUT parameter and the mass parameter had been fixed to ensure
the quaternionic property (\ref{120}) (or equivalently the
anti-selfduality of the Weyl tensor). For vanishing cosmological
constant ($\kappa=0$) one obtains the well known Taub-NUT metric, and
hence, this space represents topologically an orbifold. In fact, this
orbifold is related to the non-trivial periodicity of $\tau$, which
ensures the absence of conical singularities:
\be663
\tau \simeq \tau +{4 \pi n} \ .
\ee
By complete analogy to the Ricci-flat Taub--NUT case we can make the
orbifold action explicit.  For $\kappa = 1$ (the other case goes in
complete analogy, see also \cite{150}) the coset $\mathbb S^4 = {SO(5)
/ SO(4)}$ is defined by
\be263
(X_0)^2 + (X_1)^2 + (X_2)^2 + (X_3)^2 + (X_4)^2 = 1
\ee
with the metric
\be274 
ds^2 = (dX_0)^2 +(dX_1)^2 + (dX_2)^2 + (dX_3)^2 + (dX_4)^2 
\ee
subject to the constraint (\ref{263}).  Before imposing the
constraint, the $SO(5)$ symmetry group is manifest, but afterwards
only a subclass of these isometries are realized linearly and the
other symmetries are not manifest.  Since we are interested here in the
spherical symmetric case ($\epsilon =1$) we introduce polar coordinates in
$X_{1,2,3,4}$: $(dX_1)^2 + (dX_2)^2 + (dX_3)^2 + (dX_4)^2 = d \rho^2 +
\rho^2\, d \Omega_3$ and the constraint becomes $X_0^2 + \rho^2 = 1$.
Since $X_0 \, dX_0 = - \rho \,d \rho$ we can eliminate $X_0$ and find
for the metric
\be188
ds^2 = {d\rho^2 \over 1- {\rho^2}} + \rho^2 d \Omega_3
    = {d \bar r^2 + \bar r^2 d\Omega_3 \over 
    (1 + {\bar r^2 \over 4 })^2}
    = {dz_1 d\bar z_1 + dz_2 d \bar z_2 \over 
    \left(1 + {|z_1|^2 + |z_2|^2 \over 4 } \right)^2}
\ee
with $\rho = { \bar r \over 1 +\bar r^2/4}$.  As for the
Taub-NUT space the ${\mathbb Z}_n$ orbifold acts on the two complex
coordinates as 
\be774
z_1 \simeq e^{2\pi i \over n} z_1 \qquad , \qquad z_2 \simeq 
    e^{-{2\pi i \over n}} z_2 
\ee
which, after the change of coordinates
\be385
z_1 = r \, \cos(\theta / 2) \, e^{i {\varphi + \psi \over 2}}
\, , \quad 
z_2 = r \, \sin({\theta / 2}) \, e^{i {\varphi - \psi \over 2}}\, ,
\ee
is equivalent to (\ref{663}). Hence, this $\mathbb Z_n$ orbifold acts
on the $\mathbb S^3$ sub-space as: $\mathbb S^3 \rightarrow {\mathbb
S}^3/{\mathbb Z}_n$ in the usual way; for a discussion of orbifolds
see also \cite{550, 530, 540}.

\begin{figure} 
\begin{center}
\includegraphics[angle=0,width=130mm]{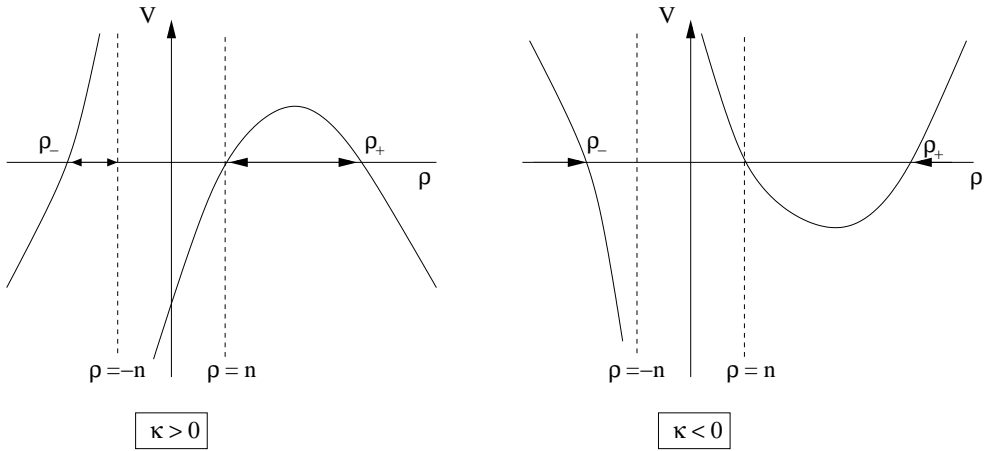}
\end{center}
\caption{A positive definite metric requires: $V>0$ as well as 
$\rho^2 > n^2$. Hence, there are two allowed coordinate regions, 
as we have indicated by the arrows.}
\label{fig2}
\end{figure}

In addition to the reduced number of isometries, given by the subgroup
that commutes with the orbifold action, there are more significant
differences to the homogeneous cases that we discussed before.  First,
in the limit $\kappa = 0$ it becomes the Taub-NUT space, which is
hyper-Kaehler and since it is still non-trivial it may be of interest
for $G_2$ manifolds as well. Second, the quaternionic space
(\ref{540}) has a curvature singularity at $\rho= - n$ as indicated
by the curvature invariant
\be550
R_{mnst} R^{mnst} = 24\Big[ \, \kappa^2 +{ 4 n^2 (\epsilon + 4 \kappa n^2)^2
	\over (\rho + n)^6} \Big] \ .
\ee
This singularity exist for any value of $u^i$ and hence also the $G_2$
manifold (\ref{510}) becomes singular, but it is shielded by the
Killing horizon where $V=0$ representing a fixed point set of $k=
\partial_\tau$, see below. Note, the zeros of $V$ are regular points
as it is obvious from the curvature invariant (\ref{550}). But at
these zeros $V$ changes its sign and therefore the metric is
well-defined only on the coordinate patch where
\[
V \, \geq \, 0 \qquad , \qquad \rho^2 - n^2 \, \geq \, 0
\]
which allows for two physical regions that are disconnected by the
curvature singularity, see figure \ref{fig2}:
\[
\rho \leq \rho_- \qquad , \qquad \rho \geq {\rm max}\{n , \rho_+\} \ .
\]
Since  the metric is invariant under $\rho \rightarrow -\rho$
and $n \rightarrow -n$ we can assume that $n>0$.

Before we discuss in the next section the Killing vectors and their
fixed point set, let us note that the solution (\ref{540})
interpolates between the two homogeneous quaternionic spaces that we
introduced in eqs.\ (\ref{520}) and (\ref{530}). To see this, let us
set $\epsilon = +1$. Obviously, if $n=0$ we get $V = 1 - \kappa
\rho^2$ and obtain the maximal symmetric spaces $\mathbb S^4$
or $EAdS_4$ in eq.\ (\ref{520}). On the other hand, if we transform
\be560
\rho = {\hat \rho \over n} + n \qquad , \qquad
\tau = 2n \psi
\ee
and take the limit
\[ 
n \rightarrow \infty \qquad {\rm keeping} \qquad \hat \rho = fix
\]
one finds the metric
\be570
ds^2 = {2 d \hat \rho^2 \over \hat \rho (1 - 4 \kappa \hat \rho)}
	+ 2 \hat \rho (1 - 4 \kappa \hat \rho) \Big[\, d\psi
	-{1 \over 2} \, {y dx -xdy \over 1 + {x^2 + y^2 \over 4}} \, \Big]^2
	+ 2 \hat \rho \, {dx^2 + dy^2 \over (1 + {x^2 +y^2 \over 4})^2} \ .
\ee
Finally, the transformation
\[
\hat \rho = { r^2 \over 4 (1+ \kappa \, r^2)} \qquad , \qquad
	x + {\rm i} \, y = 2 \, \tanh{\theta \over 2} \, e^{{\rm i} 
	\varphi} \ .
\]
brings us to the complex projective space in eq.\  (\ref{530}). 

Since negative $n$ are equivalent to positive (after $\rho \rightarrow
- \rho$), the two homogenous spaces appear at the endpoints of the
parameter space of $n$ and in both limits the curvature singularity
disappears and the spaces become smooth.  Let us also note, that in
the limit of vanishing cosmological constant the space (\ref{540})
is the standard Taub-NUT space, whereas the space
(\ref{570}) which appears in the large $n$ limit becomes for $\kappa =
0$ the Eguchi-Hanson space.

%%%%%%%%%%%%%%%%%%%%%%%%%%%%%%%%%%%%%%%%%%%%%%%%%%%%%

\section{Isometries and IIA description}

%%%%%%%%%%%%%%%%%%%%%%%%%%%%%%%%%%%%%%%%%%%%%%%%%%

Having the metric of the manifold $X_7$ we can consider M-theory on
$M_4 \times X_7$ and if $X_7$ has at least one isometry the
dimensional reduction will give a IIA description.  Especially
interesting are Killing vector fields with a fixed point set $L$ of
co-dimension four implying that $L$ extends in 6+1 dimensions, and
hence, become D6-branes upon compactification \cite{170}.  But before
we come to the fixed point set let us discuss the Killing vectors.

%%%%%%%%%%%%%%%%%%%%%%%%%%%%%%%%%%%%%%%%

\subsection{Isometries}

%%%%%%%%%%%%%%%%%%%%%%%%%%%%%%%%%%%%%%%%

An isometry is related to the existence of a Killing vector field $k$
satisfying the equation $D_{(m} k_{n)} = 0$. In a given coordinate
patch one can introduce proper coordinates so that the Killing vector
field becomes $k = \partial_\chi$ and the ($\chi$-independent) metric  
reads
\be600
ds^2 = e^{2 \phi(x)} \Big[ d\chi + \omega_m(x) dx^m \Big]^2 + \hat g_{mn}(x) 
	dx^m dx^n \ .
\ee
In these (local) coordinates the Killing vector field becomes $k_M =
e^{2 \phi} \{1, \omega_m \}$ and due to the Killing property one find
$D_M k_N = D_{[M} k_{N]}= \partial_{[M} k_{N]}$. Obviously any
covariantly constant vector field is also a Killing vector with the
consequence that $\phi$ and $\omega_m$ are trivial (i.e.\ $d\omega
=0$). This in turn implies that the space factorizes and the holonomy
is reduced. Therefore, the holonomy in our case is equal to $G_2$ iff
there are no covariantly constant (Killing) vector fields.  The
absence of such Killing vectors for 7-manifolds given as ${\mathbb
R}^3$ bundles over ${\mathbb S^4}$ and ${\mathbb C \mathbb P}^2$ was
shown shown in \cite{110, 100}. Since our case represents an
interpolation between these two spaces, where only a subclass of
isometries survive, we are still dealing with a manifold where the
holonomy is equal to $G_2$. 

To be more concrete, the isometries of $X_7$ with the metric
(\ref{510}) are given by the isometries of the quaternionic space
(\ref{540}) and in addition $SU(2)$ transformations of the coordinates
$u^i$. The quaternionic space has four Killing vectors \cite{450}
which can be written as \cite{150}
\be610
\ba{l}
k_1 = n y \, \partial_\tau - (1 + \epsilon \, {x^2 -y^2 \over 4})\, 
	\partial_x - \epsilon \, {xy \over 2} \, \partial_y \ , \\[3mm]
k_2 = - n x \, \partial_\tau - \epsilon \, {xy \over 2} \, \partial_x 
	- (1 - \epsilon \, {x^2 -y^2 \over 4})\, \partial_y \ , \\[3mm]
k_3 = 2n \, \partial_\tau + \epsilon \, y \, \partial_x -
	\epsilon \, x \, \partial_y
	 \quad , \qquad k_4 = \partial_\tau
\ea
\ee
and fulfill the algebra 
\be620 
\big[\, k_i , k_4 \, \big] =0 \quad , \qquad \big[\, k_i , k_j \, \big]
	= f_{ijl} \, k_l
\ee
with $f_{123} = 1$, $f_{231} = f_{312} = \epsilon$. For the spherical
symmetric case $\epsilon = 1$ we get thus the symmetry $SU(2) \times
U(1)$, where the $U(1)$ corresponds to the orbifold action (\ref{774})
and the $SU(2)$ is the subgroup of $SU(3)$ or $SU(2,1)$ that commutes
with the orbifold.  For the hyperbolic case ($\epsilon = -1$) one
obtains the non-compact analog, namely the algebra $SL(2,R) \times
U(1)$. For $\epsilon = 0$ two Killing vectors become equivalent ($k_3
\sim k_4$) and we should take a different parameterization, namely
\be622
k_1 = n y \, \partial_\tau -  \partial_x \ , \quad
k_2 = - n x \, \partial_\tau  - \partial_y \ , \quad
k_3 = y \, \partial_x - x \, \partial_y
	 \ , \quad k_4 = \partial_\tau \ .
\ee
They satisfy the algebra
\be624
\big[\, k_i , k_4 \, \big] =0
 \  , \quad \big[\, k_1 , k_2 \, \big]	= 2 n k_4 
 \  , \quad \big[\, k_2 , k_3 \, \big]	= k_1 
 \  , \quad \big[\, k_3 , k_1 \, \big]	= k_2 
\ee
which is not not anymore a direct product of the two Lie algebras.

{From} our discussion above it is clear that a covariantly constant
Killing vector would commute with the other Killing vectors.  But the
only commuting Killing vector $k_4 = \partial_\tau$ is not covariantly
constant, due to the non-trivial $U(1)$ fibration in (\ref{540}).
Therefore the space cannot be factorized and the holonomy is equal to
$G_2$, which we infered already from the relation to the other known
$G_2$ manifolds.

%%%%%%%%%%%%%%%%%%%%%%%%%%%%%%%%%%%%%%%%%%%%%%

\subsection{Fixed point set}

%%%%%%%%%%%%%%%%%%%%%%%%%%%%%%%%%%%%%%%%%%%%%%

For a given Killing vector $k$, the hypersurface $L$ defined by $|k|^2
= 0$ is the fixed point set and $L$ degenerates if the surface gravity
vanishes, i.e.\ if $|Dk|^2 = 0$ at $|k|^2=0$; see e.g.\ \cite{460,
470}.  For $\epsilon \neq 0$ the fixed points of all Killing vectors
are non-degenerate implying a periodic identification along the
Killing direction to ensure the absence of conical singularities. This
fact is well known from {\em non-extreme} black holes where the event
horizon is a non-degenerate fixed point set for a timelike Killing
vector. In the case here, the compactness of the Killing direction
ensures that the Kaluza-Klein gauge group becomes $U(1)$ and that the
D6-branes, which are identified as the fixed point set $L$ \cite{170},
are at finite geodesic distance.

In order to identify the D6-branes we need a Killing vector that has a
co-dimension four fixed point set $L$, or in other words, $L$ extends
over three coordinates in $X_7$ which, in addition to $M_4$, become
the world volume of the D6-branes. We should therefore not consider
the Killing vectors related rotations of the $u^i$ coordinates.  They
have a fixed point set at $|u| = 0$ which is a point in the 3-d
$u^i$-space and hence a co-dimension three fixed point set.  On the
other hand, the fixed points of the quaternionic Killing vectors can be
NUTs or bolts, depending on the rank of the 2-from $dk$; see
\cite{460}.  The NUTs are points on the quaternionic space and since
the $SU(2)$ connection $A^i$ becomes trivial, these NUTs
correspond to isotropic D6-branes.  On the other hand, if $L$
is a bolt on the quaternionic space it has co-dimension two and since
at least one $SU(2)$ connection remains non-trivial, there is no
isotropic brane interpretation. After dimensional reduction
we obtain a supergravity solution which is singular at the fixed point
set $L$ and if $L$ is a NUT the singularity can be identified with the
location of the D6-brane, but if $L$ is a bolt there is no clear
interpretation. It may be related to some deformation of a given
D6-brane, since, as we will see, the bolts are connected with the
NUTs for specific Killing vectors.

A generic Killing vector of the quaternionic space, can be introduced
as a linear combination of the four Killing vectors in
(\ref{610})
\be630
\ba{rcl}
k &=& \alpha_1 k_1 + \alpha_2 k_2 + \alpha_3 k_3 + \alpha_4 k_4 \\[3mm]
	&=& \  \Big[ \, ny \, \alpha_1 - n x \, \alpha_2 + 2n \, \alpha_3 + 
	\alpha_4 \, \Big] \, \partial_{\tau}  \\[3mm] 
&&	- \Big[ \, (1 + \epsilon \, {x^2 - y^2 \over 4}) 
	\, \alpha_1 + \epsilon\, {xy \over 2} \, \alpha_2 + 
	\epsilon\, y \, \alpha_3 \, \Big] \, \partial_x  \\[3mm]
&&	- \Big[\,  \epsilon\, {xy \over 2} \, \alpha_1 + (1 - 
	\epsilon\, {x^2 -y^2 \over 4}) \,
	\alpha_2 + \epsilon\, x \, \alpha_3 \, \Big] \, \partial_y \ .
\ea
\ee
We will now investigate the fixed point set of this Killing vector
following the discussion in \cite{150}.  Neglecting zeros of
$V(\rho)$, zeros $|k|^2$ are given by zeros of all components
$k^m$. Since the Killing vector depends only on $x$ and $y$ the
equations $k^m = 0$ are not solvable for a generic choice of
$\alpha_m$. We can solve, however, $k^x = k^y=0$ and find as solution
\cite{150}
\be640
x=x_{\pm} ={2 \alpha_2 \over \alpha_1^2 + \alpha_2^2} \Big( \, \alpha_3
	\pm |\alpha| \, \Big) \quad ,\qquad
y = y_{\pm} = - {2 \alpha_1  \over \alpha_1^2 + \alpha_2^2} \Big( \, \alpha_3
	\pm |\alpha| \, \Big)
\ee
where $|\alpha|^2 = \alpha_1^2 + \alpha_2^2 + \alpha_3^2$.  If one
inserts these values into $k^\tau = 0$ one obtains a constraint on
$\alpha_4$, but one would obtain bolts since $\rho$ and $\tau$ are
arbitrary. In order to find a NUT we keep $\alpha_4$ arbitrary and set
instead
\be642
V(\rho) = 0 \quad  \quad {\rm or}: \quad\rho = (n , \rho_\pm)
\ee
which ensures that $g_{\tau\tau} k^\tau k^\tau = 0$ and represent
points in the $(\rho , \tau)$ space\footnote{Notice, $V=0$ is a
conical singularity which is resolved by the proper periodicity in
$\tau$.}.  The point $\rho = n$ is special, because without fixing
$x,y$ we find $|k|^2 \equiv 0$.

Therefore, we have on the quaternionic space the following NUT fixed
points of the Killing vector $k$
\be650
\ba{l}
\kappa>0: \ (\rho, x,y) = \big\{\, (n, x, y) \, , \,  
	(\rho_- , x_{\pm} , y_{\pm}) \, (\rho_+ , x_{\pm} , 
	y_{\pm})_{{\rm if}\  \epsilon>0} \,  \big\}
	\\[2mm]
\kappa<0: \ (\rho, x,y) = \big\{\, ({\rm max}[n, \rho_+], x_\pm , y_\pm ) \, , 
	 \, (\rho_- , x_{\pm} , y_{\pm}) \, \big\}
\ea
\ee
with $\rho_\pm = - n \pm \sqrt{4n^2 + {\epsilon \over \kappa}}$.

The dimensional reduction along the Killing vector (\ref{630}) will
yield a bound state of D6-branes, located on the quaternionic space at
these fixed points.  The number depends on the choice of parameters,
but can be at most five (for $\epsilon, \kappa > 0$) where two of them
(at $\rho =\rho_-$) are disconnected from the other three (at $\rho =
n$ and $\rho=\rho_+$) by the curvature singularity at $\rho = -n$.
Recall, there are different (physical) regions for $\rho$ (see figure
\ref{fig2}) and hence some fixed points may not have a physical
sensible interpretation.  The correct identification of the D6-branes
is, however, a subtle point. E.g.\ because $x_+ x_- + y_+ y_- = -4$
the two points ($x_+, y_+$) and ($x_-, y_-$) are antipodal points in
the ($x,y$)-space (i.e.\ for $\epsilon=1$ on the $\mathbb S^2$) and
one may be tempted to identify the two branes. But note, since the
Killing vector is zero at both points and non-singular in between, the
eigenvalue of $dk$ have to be different at both fixed points, which
implies that the 6-brane charge or tension will be different.  It
would be interesting to see whether our model corresponds to one of
the supersymmetric intersecting brane worlds or the relation to the
non-supersymmetric ones; see e.g.\ \cite{590, 600} and refs.\ therein.

A subtle question concerns also the dilaton. In adapted coordinates
one can write the 7-metric as in eq.\ (\ref{600}), where $x_{11}$
correspond to $\chi$ and $\phi$ is proportional to the dilaton.
Unfortunately, $e^{2 \phi} \sim |k|^2$, which becomes the string
coupling constant, is not bounded for our solution. While for $\kappa
> 0$ the range of $\rho$ is bounded, see figure \ref{fig2}, but $|u|$
is unbounded, see eq.\ (\ref{260}). On the other hand for a
non-compact quaternionic space, with $\kappa \leq 0$, is situation is
opposite.  Therefore, in any case there are asymptotic regions where
$|k|^2 \sim e^{2\phi}$ blows up and one should look for similar
deformation as the ones discussed in \cite{270} to obtain a finite
string coupling constant.

%%%%%%%%%%%%%%%%%%%%%%%%%%%%%%%%%%%%%%%%%%%%%%%%%%%%%%%%%%%%%

\section{Outlook: Membrane instantons and new spin(7) metric}

%%%%%%%%%%%%%%%%%%%%%%%%%%%%%%%%%%%%%%%%%%%%%%%%%%%%%%%%%%%%

We mentioned already a few  aspects that deserve further investigations,
but let us add another two.

\bigskip

\noindent
{\em A. Resolution of the curvature singularity by Membrane instantons}

\medskip

\noindent
One genuine feature of our $G_2$ manifold is the appearance of a
curvature singularity; otherwise the continues deformation of two
topological different spaces (like $\mathbb S^4$ and $\mathbb C
\mathbb P^2$) would be difficult to understand. Although for
$\kappa<0$ the singularity is inside an unphysical coordinate range
(with timelike coordinates), the validity of the supergravity solution
is a subtle question.  One may therefore look for possibilities to
resolve the singularity. An interesting possibility could be to turn
on appropriate fluxes along the lines discussed in \cite{210,
220}. But non-trivial fluxes play an important role also in a somewhat
different aspect.  Namely the curvature singularity in eq.\
(\ref{550}) disappears if $\epsilon + 4 \kappa n^2 = 0$ (or $\rho_- =
\rho_+$) and since $\kappa$ is the constant curvature of the 4-d
quaternionic space, this value can effectively be changed by turning
on 4-form fluxes. In fact, this possibility has been discussed as
``Neutralization of the cosmological constant'' and corresponds to
take into account membrane instanton effects \cite{380, 390}.  What
will happen with the quaternionic space in the limit $\rho_+ =
\rho_-$?  Setting $\epsilon = 1$ and $\kappa =-1$ so that $n=1/2$ we
transform $\rho = {1 \over 2} \cosh R$ and the metric becomes
\be750
ds^2 = dR^2 + \sinh^2 R \, d\Omega_3 
\ee
which is one parameterization of Euclidean $AdS_4$ space.  This metric
covers only the region $\rho > n$, but for $\rho < -n$ one obtains the
same metric so that both physical regions are decoupled and the region
with the curvature singularity disappeared and moreover all fixed
points are joined at $\rho = \pm n$ which corresponds to the point
$R=0$ where the metric becomes flat. But note, the 4-form flux will
not only change the effective value of the curvature of the
quaternionic space, it will also cause a back reaction on the 7-metric
which requires a detailed analysis. Needless to say, that it would be
also interesting to calculate the superpotential caused by these
4-form fluxes \cite{260, 340, 250}.

There is also another effect that may weaken or resolve the
curvature singularity, namely to change effectively $n$ by a
multi-center solution (as for Taub-NUT or the Eguchi-Hanson space).
It is unclear whether these multi-center solution for quaternionic
spaces exist, but one could try to construct them similar to the
multi-center supergravity solutions of \cite{440, 430}.

\bigskip

\noindent
{\em B. New Spin(7) metrics}

\medskip

\noindent
We have considered here only generalizations of known 7-manifolds with
$G_2$ holonomy, but analogous calculations will also yield more general
8-manifolds with Spin(7) holonomy. If one follows again the work by
Bryant and Salamon, the corresponding metric of the 8-manifold becomes
\be700
ds^2 = f^2 \alpha \bar \alpha + g^2 e^m e^m
\ee
where $e^m e^m$ is again the metric of a quaternionic space and
$\alpha = du - u A$, where we used the quaternionic notation $u = u^0
+ i u^1 +j u^2 + k u^3$ and $A = i A^1 + j A^2 + k A^3$ is the $SU(2)$
connection of the quaternionic space. The function $f = f(|u|)$ and $g
= g(|u|)$ are again the same as in \cite{110} and new fixed points are
again encoded by the quaternionic space. Also, one may turn on fluxes
and calculate the superpotential or replace the $\mathbb R^4$, spanned
by the coordinates $u^m$, by a hyper Kaehler space and finds a IIA
description of 6-branes wrapping the quaternionic space. Some related
work in these directions has been done e.g.\ in
\cite{691,560,190,570,580}.

%%%%%%%%%%%%%%%%%%%%%%%%%%%%%%%%%%%%%%%%%%%%%

\vspace{10mm}

\noindent
{\bf Acknowledgments}

\medskip

\noindent
We would like to thank R.\ Blumenhagen, G.\ Dall'Agata, S.\ Gukov and
A.\ Klemm for helpful discussions. This work is supported by a
Heisenberg grant of the DFG.

%%%%%%%%%%%%%%%%%%%%%%%%%%%%%%%%%%%%%%%%%%%%%%%%%%%%%%

% ---- Bibliography ----

% \nocite{*} %this uses *everything* in the .bib file
% \bibliography{g2-metric} %or whatever your .bib file is

\begin{thebibliography}{10}

\bibitem{420}
M.~J. Duff, B.~E.~W. Nilsson, and C.~N. Pope, ``Kaluza-{K}lein supergravity,''
  {\em Phys. Rept.} {\bf 130} (1986) 1--142.

\bibitem{370}
G.~Papadopoulos and P.~K. Townsend, ``Compactification of {D} = 11 supergravity
  on spaces of exceptional holonomy,'' {\em Phys. Lett.} {\bf B357} (1995)
  300--306,
  \href{http://xxx.lanl.gov/abs/http://arXiv.org/abs/hep-th/9506150}{{\tt
  http://arXiv.org/abs/hep-th/9506150}}.

\bibitem{110}
R.~L. Bryant and S.~M. Salamon, ``On the construction of some complete metrics
  with exceptional holonomy,'' {\em Duke Math. Journal} {\bf 58} (1989) 829.

\bibitem{100}
G.~W. Gibbons, D.~N. Page, and C.~N. Pope, ``Einstein metrics on ${S^3}$,
  ${R^3}$ and ${R^4}$ bundles,'' {\em Commun. Math. Phys.} {\bf 127} (1990)
  529.

\bibitem{640}
E.~Witten, ``Fermion quantum numbers in {K}aluza-{K}lein theory,''. in
  *Appelquist, T. (ed.) et al.: Modern Kaluza-Klein theories*, 438-511. (in
  *Shelter Island 1983, Proccedings, Quantum field theory and the fundamental
  problems of physics*, 227-277) and preprint - Witten, E. (83,rec.Jan.84) 78
  P. (see book index).

\bibitem{650}
B.~S. Acharya, ``On realising {N} = 1 super {Yang-Mills} in {M} theory,''
  \href{http://xxx.lanl.gov/abs/http://arXiv.org/abs/hep-th/0011089}{{\tt
  http://arXiv.org/abs/hep-th/0011089}}.

\bibitem{590}
M.~Cvetic, G.~Shiu, and A.~M. Uranga, ``Chiral four-dimensional {N} = 1
  supersymmetric type {IIA} orientifolds from intersecting {D6}-branes,'' {\em
  Nucl. Phys.} {\bf B615} (2001) 3--32,
  \href{http://xxx.lanl.gov/abs/http://arXiv.org/abs/hep-th/0107166}{{\tt
  http://arXiv.org/abs/hep-th/0107166}}.

\bibitem{350}
B.~Acharya and E.~Witten, ``Chiral fermions from manifolds of {G(2)}
  holonomy,''
  \href{http://xxx.lanl.gov/abs/http://arXiv.org/abs/hep-th/0109152}{{\tt
  http://arXiv.org/abs/hep-th/0109152}}.

\bibitem{360}
M.~Atiyah and E.~Witten, ``M-theory dynamics on a manifold of {G(2)}
  holonomy,''
  \href{http://xxx.lanl.gov/abs/http://arXiv.org/abs/hep-th/0107177}{{\tt
  http://arXiv.org/abs/hep-th/0107177}}.

\bibitem{550}
M.~Atiyah, J.~Maldacena, and C.~Vafa, ``An {M}-theory flop as a large {N}
  duality,'' {\em J. Math. Phys.} {\bf 42} (2001) 3209--3220,
  \href{http://xxx.lanl.gov/abs/http://arXiv.org/abs/hep-th/0011256}{{\tt
  http://arXiv.org/abs/hep-th/0011256}}.

\bibitem{240}
M.~Cvetic, H.~Lu, and C.~N. Pope, ``Massless 3-branes in {M}-theory,'' {\em
  Nucl. Phys.} {\bf B613} (2001) 167--188,
  \href{http://xxx.lanl.gov/abs/http://arXiv.org/abs/hep-th/0105096}{{\tt
  http://arXiv.org/abs/hep-th/0105096}}.

\bibitem{270}
A.~Brandhuber, J.~Gomis, S.~S. Gubser, and S.~Gukov, ``Gauge theory at large
  {N} and new {G(2)} holonomy metrics,'' {\em Nucl. Phys.} {\bf B611} (2001)
  179--204,
  \href{http://xxx.lanl.gov/abs/http://arXiv.org/abs/hep-th/0106034}{{\tt
  http://arXiv.org/abs/hep-th/0106034}}.

\bibitem{230}
M.~Cvetic, G.~W. Gibbons, H.~Lu, and C.~N. Pope, ``Cohomogeneity one manifolds
  of {Spin(7)} and {G(2)} holonomy,''
  \href{http://xxx.lanl.gov/abs/http://arXiv.org/abs/hep-th/0108245}{{\tt
  http://arXiv.org/abs/hep-th/0108245}}.

\bibitem{250}
S.~Gukov and D.~Tong, ``D-brane probes of special holonomy manifolds, and
  dynamics of {N} = 1 three-dimensional gauge theories,''
  \href{http://xxx.lanl.gov/abs/http://arXiv.org/abs/hep-th/0202126}{{\tt
  http://arXiv.org/abs/hep-th/0202126}}.

\bibitem{530}
T.~Friedmann, ``On the quantum moduli space of {M} theory compactifications,''
  \href{http://xxx.lanl.gov/abs/http://arXiv.org/abs/hep-th/0203256}{{\tt
  http://arXiv.org/abs/hep-th/0203256}}.

\bibitem{540}
H.~Ita, Y.~Oz, and T.~Sakai, ``Comments on {M} theory dynamics on {G(2)}
  holonomy manifolds,''
  \href{http://xxx.lanl.gov/abs/http://arXiv.org/abs/hep-th/0203052}{{\tt
  http://arXiv.org/abs/hep-th/0203052}}.

\bibitem{660}
R.~Blumenhagen and V.~Braun, ``Superconformal field theories for compact {G(2)}
  manifolds,'' {\em JHEP} {\bf 12} (2001) 006,
  \href{http://xxx.lanl.gov/abs/http://arXiv.org/abs/hep-th/0110232}{{\tt
  http://arXiv.org/abs/hep-th/0110232}}.

\bibitem{720}
A.~Brandhuber, ``G(2) holonomy spaces from invariant three-forms,''
  \href{http://xxx.lanl.gov/abs/http://arXiv.org/abs/hep-th/0112113}{{\tt
  http://arXiv.org/abs/hep-th/0112113}}.

\bibitem{320}
J.~Wolf, ``Complex homogenous contact manifolds and quaternionic symmetric
  spaces,'' {\em J. Math. Mech.} {\bf 14} (1965) 1033.

\bibitem{620}
J.~Bagger and E.~Witten, ``Matter couplings in {N=2} supergravity,'' {\em Nucl.
  Phys.} {\bf B222} (1983) 1.

\bibitem{610}
B.~de~Wit, P.~G. Lauwers, and A.~Van~Proeyen, ``Lagrangians of {N=2}
  supergravity - matter systems,'' {\em Nucl. Phys.} {\bf B255} (1985) 569.

\bibitem{630}
B.~de~Wit, M.~Rocek, and S.~Vandoren, ``Hypermultiplets, hyperkaehler cones and
  quaternion-{Kaehler} geometry,'' {\em JHEP} {\bf 02} (2001) 039,
  \href{http://xxx.lanl.gov/abs/http://arXiv.org/abs/hep-th/0101161}{{\tt
  http://arXiv.org/abs/hep-th/0101161}}.

\bibitem{490}
R.~D'Auria and S.~Ferrara, ``On fermion masses, gradient flows and potential in
  supersymmetric theories,'' {\em JHEP} {\bf 05} (2001) 034,
  \href{http://xxx.lanl.gov/abs/http://arXiv.org/abs/hep-th/0103153}{{\tt
  http://arXiv.org/abs/hep-th/0103153}}.

\bibitem{670}
R.~Kallosh and A.~D. Linde, ``Supersymmetry and the brane world,'' {\em JHEP}
  {\bf 02} (2000) 005,
  \href{http://xxx.lanl.gov/abs/http://arXiv.org/abs/hep-th/0001071}{{\tt
  http://arXiv.org/abs/hep-th/0001071}}.

\bibitem{680}
K.~Behrndt and M.~Cvetic, ``Anti-de sitter vacua of gauged supergravities with
  8 supercharges,'' {\em Phys. Rev.} {\bf D61} (2000) 101901,
  \href{http://xxx.lanl.gov/abs/http://arXiv.org/abs/hep-th/0001159}{{\tt
  http://arXiv.org/abs/hep-th/0001159}}.

\bibitem{150}
K.~Behrndt and G.~Dall'Agata, ``Vacua of {N} = 2 gauged supergravity derived
  from non-homogeneous quaternionic spaces,''
  \href{http://xxx.lanl.gov/abs/http://arXiv.org/abs/hep-th/0112136}{{\tt
  http://arXiv.org/abs/hep-th/0112136}}.

\bibitem{170}
P.~K. Townsend, ``The eleven-dimensional supermembrane revisited,'' {\em Phys.
  Lett.} {\bf B350} (1995) 184--187,
  \href{http://xxx.lanl.gov/abs/http://arXiv.org/abs/hep-th/9501068}{{\tt
  http://arXiv.org/abs/hep-th/9501068}}.

\bibitem{500}
D.~V. Alekseevsky, V.~Cortes, C.~Devchand, and A.~Van~Proeyen, ``Flows on
  quaternionic-{Kaehler} and very special real manifolds,''
  \href{http://xxx.lanl.gov/abs/http://arXiv.org/abs/hep-th/0109094}{{\tt
  http://arXiv.org/abs/hep-th/0109094}}.

\bibitem{130}
B.~de~Wit and H.~Nicolai, ``The parallelizing {S(7)} torsion in gauged {N=8}
  supergravity,'' {\em Nucl. Phys.} {\bf B231} (1984) 506.

\bibitem{120}
A.~Bilal, J.-P. Derendinger, and K.~Sfetsos, ``{(Weak)} {G(2)} holonomy from
  self-duality, flux and supersymmetry,''
  \href{http://xxx.lanl.gov/abs/http://arXiv.org/abs/hep-th/0111274}{{\tt
  http://arXiv.org/abs/hep-th/0111274}}.

\bibitem{140}
M.~Gunaydin and F.~Gursey, ``Quark structure and octonions,'' {\em J. Math.
  Phys.} {\bf 14} (1973) 1651--1667.

\bibitem{400}
S.~Fubini and H.~Nicolai, ``The octonionic instanton,'' {\em Phys. Lett.} {\bf
  B155} (1985) 369.

\bibitem{410}
B.~de~Wit, H.~Nicolai, and N.~P. Warner, ``The embedding of gauged {N=8}
  supergravity into {D} = 11 supergravity,'' {\em Nucl. Phys.} {\bf B255}
  (1985) 29.

\bibitem{700}
E.~G. Floratos and A.~Kehagias, ``Eight-dimensional self-dual spaces,'' {\em
  Phys. Lett.} {\bf B427} (1998) 283--290,
  \href{http://xxx.lanl.gov/abs/http://arXiv.org/abs/hep-th/9802107}{{\tt
  http://arXiv.org/abs/hep-th/9802107}}.

\bibitem{710}
B.~S. Acharya and M.~O'Loughlin, ``Self-duality in {D $\leq$} 8-dimensional
  {Euclidean} gravity,'' {\em Phys. Rev.} {\bf D55} (1997) 4521--4524,
  \href{http://xxx.lanl.gov/abs/http://arXiv.org/abs/hep-th/9612182}{{\tt
  http://arXiv.org/abs/hep-th/9612182}}.

\bibitem{330}
D.~Alekseevsky, ``Classification of quaternionic spaces with a transitive
  solvable group of motions,'' {\em Math USSR} {\bf Izv. 9} (1975) 297.

\bibitem{690}
K.~Behrndt and M.~Cvetic, ``Gauging of {N=2} supergravity hypermultiplet and
  novel renormalization group flows,'' {\em Nucl. Phys.} {\bf B609} (2001)
  183--192,
  \href{http://xxx.lanl.gov/abs/http://arXiv.org/abs/hep-th/0101007}{{\tt
  http://arXiv.org/abs/hep-th/0101007}}.

\bibitem{510}
P.-Y. Casteill, E.~Ivanov, and G.~Valent, ``Quaternionic extension of the
  double {Taub-NUT} metric,'' {\em Phys. Lett.} {\bf B508} (2001) 354--364,
  \href{http://xxx.lanl.gov/abs/http://arXiv.org/abs/hep-th/0104078}{{\tt
  http://arXiv.org/abs/hep-th/0104078}}.

\bibitem{520}
P.-Y. Casteill, E.~Ivanov, and G.~Valent, ``{U(1) x U(1)} quaternionic metrics
  from harmonic superspace,'' {\em Nucl. Phys.} {\bf B627} (2002) 403--444,
  \href{http://xxx.lanl.gov/abs/http://arXiv.org/abs/hep-th/0110280}{{\tt
  http://arXiv.org/abs/hep-th/0110280}}.

\bibitem{450}
D.~Kramer, E.~Herlt, M.~MacCallum, and H.~Stephani, ``Exact solutions of
  {E}instein's field equations,'' {\em Cambridge University Press} (1979).

\bibitem{460}
G.~W. Gibbons and S.~W. Hawking, ``Classification of gravitational instanton
  symmetries,'' {\em Commun. Math. Phys.} {\bf 66} (1979) 291.

\bibitem{470}
P.~K. Townsend, ``Black holes,''
  \href{http://xxx.lanl.gov/abs/http://arXiv.org/abs/gr-qc/9707012}{{\tt
  http://arXiv.org/abs/gr-qc/9707012}}.

\bibitem{600}
R.~Blumenhagen, B.~Kors, D.~Lust, and T.~Ott, ``The standard model from stable
  intersecting brane world orbifolds,'' {\em Nucl. Phys.} {\bf B616} (2001)
  3--33,
  \href{http://xxx.lanl.gov/abs/http://arXiv.org/abs/hep-th/0107138}{{\tt
  http://arXiv.org/abs/hep-th/0107138}}.

\bibitem{210}
M.~Cvetic, G.~W. Gibbons, H.~Lu, and C.~N. Pope, ``Ricci-flat metrics, harmonic
  forms and brane resolutions,''
  \href{http://xxx.lanl.gov/abs/http://arXiv.org/abs/hep-th/0012011}{{\tt
  http://arXiv.org/abs/hep-th/0012011}}.

\bibitem{220}
M.~Cvetic, H.~Lu, and C.~N. Pope, ``Brane resolution through transgression,''
  {\em Nucl. Phys.} {\bf B600} (2001) 103--132,
  \href{http://xxx.lanl.gov/abs/http://arXiv.org/abs/hep-th/0011023}{{\tt
  http://arXiv.org/abs/hep-th/0011023}}.

\bibitem{380}
J.~D. Brown and C.~Teitelboim, ``Neutralization of the cosmological constant by
  membrane creation,'' {\em Nucl. Phys.} {\bf B297} (1988) 787--836.

\bibitem{390}
R.~Bousso and J.~Polchinski, ``Quantization of four-form fluxes and dynamical
  neutralization of the cosmological constant,'' {\em JHEP} {\bf 06} (2000)
  006, \href{http://xxx.lanl.gov/abs/http://arXiv.org/abs/hep-th/0004134}{{\tt
  http://arXiv.org/abs/hep-th/0004134}}.

\bibitem{260}
B.~Acharya, X.~de~la Ossa, and S.~Gukov, ``G-flux, supersymmetry and {Spin(7)}
  manifolds,''
  \href{http://xxx.lanl.gov/abs/http://arXiv.org/abs/hep-th/0201227}{{\tt
  http://arXiv.org/abs/hep-th/0201227}}.

\bibitem{340}
C.~Beasley and E.~Witten, ``A note on fluxes and superpotentials in {M-theory}
  compactifications on manifolds of {G(2)} holonomy,''
  \href{http://xxx.lanl.gov/abs/http://arXiv.org/abs/hep-th/0203061}{{\tt
  http://arXiv.org/abs/hep-th/0203061}}.

\bibitem{440}
L.~A.~J. London, ``Arbitrary dimensional cosmological multi - black holes,''
  {\em Nucl. Phys.} {\bf B434} (1995) 709--735.

\bibitem{430}
J.~T. Liu and W.~A. Sabra, ``Multi-centered black holes in gauged {D} = 5
  supergravity,'' {\em Phys. Lett.} {\bf B498} (2001) 123--130,
  \href{http://xxx.lanl.gov/abs/http://arXiv.org/abs/hep-th/0010025}{{\tt
  http://arXiv.org/abs/hep-th/0010025}}.

\bibitem{691}
M.~Cvetic, G.~W. Gibbons, H.~Lu, and C.~N. Pope, ``New complete non-compact
  {Spin(7)} manifolds,'' {\em Nucl. Phys.} {\bf B620} (2002) 29--54,
  \href{http://xxx.lanl.gov/abs/http://arXiv.org/abs/hep-th/0103155}{{\tt
  http://arXiv.org/abs/hep-th/0103155}}.

\bibitem{560}
R.~Hernandez, ``Branes wrapped on coassociative cycles,'' {\em Phys. Lett.}
  {\bf B521} (2001) 371--375,
  \href{http://xxx.lanl.gov/abs/http://arXiv.org/abs/hep-th/0106055}{{\tt
  http://arXiv.org/abs/hep-th/0106055}}.

\bibitem{190}
G.~Curio, B.~Kors, and D.~Lust, ``Fluxes and branes in type {II} vacua and
  {M-theory} geometry with {G(2)} and {Spin(7)} holonomy,''
  \href{http://xxx.lanl.gov/abs/http://arXiv.org/abs/hep-th/0111165}{{\tt
  http://arXiv.org/abs/hep-th/0111165}}.

\bibitem{570}
B.~Acharya, X.~de~la Ossa, and S.~Gukov, ``G-flux, supersymmetry and spin(7)
  manifolds,''
  \href{http://xxx.lanl.gov/abs/http://arXiv.org/abs/hep-th/0201227}{{\tt
  http://arXiv.org/abs/hep-th/0201227}}.

\bibitem{580}
R.~Hernandez and K.~Sfetsos, ``An eight-dimensional approach to {G(2)}
  manifolds,''
  \href{http://xxx.lanl.gov/abs/http://arXiv.org/abs/hep-th/0202135}{{\tt
  http://arXiv.org/abs/hep-th/0202135}}.

\end{thebibliography}
% \bibliographystyle{utphys} %if you use utphys.bst

\providecommand{\href}[2]{#2}\begingroup\raggedright\endgroup

\end{document}